\documentclass[structabstract]{aa}

\usepackage{natbib}                                 
\usepackage{graphicx}
\usepackage{txfonts}
\def\lsim{\lower.5ex\hbox{$\; \buildrel < \over \sim \;$}}
\def\gsim{\lower.5ex\hbox{$\; \buildrel > \over \sim \;$}}
\begin{document}
%
\title{Nuclear Interaction Gamma-Ray Lines \\ from the Galactic Center Region}

\author{V. A. Dogiel\inst{1,2}, V. Tatischeff\inst{3},
K. S. Cheng\inst{4}, D. O. Chernyshov\inst{1,2,4,5},
C. M. Ko\inst{2}\thanks{cmko@astro.ncu.edu.tw}, \and W. H. Ip\inst{2}
}

\institute{I.E.Tamm Theoretical Physics Division of P.N.Lebedev
Institute, Leninskii pr, 53, 119991 Moscow, Russia
\and
Institute of Astronomy, National Central University, Jhongli 320, Taiwan
\and
Centre de Spectrom\'etrie Nucl\'eaire et de
Spectrom\'etrie de Masse, IN2P3-CNRS and Univ Paris-Sud, F-91405
Orsay Campus, France \and Department of Physics, University of
Hong Kong, Pokfulam Road, Hong Kong, China \and
Moscow Institute of Physics and Technology, Institutskii lane, 141700 
Moscow Region, Dolgoprudnii, Russia}

\date{Received...; accepted...}


\abstract
{}
{The accretion of stars onto the central supermassive black hole at
the center of the Milky Way is predicted to generate large fluxes of
subrelativistic ions in the Galactic center region. We analyze the intensity,
shape and spatial distribution of de-excitation gamma-ray lines produced by
nuclear interactions of these energetic particles with the ambient medium. }
{We first estimate the amount and mean kinetic energy of particles
released from the central black hole during star disruption. We
then calculate from a kinetic equation the energy and spatial distributions of
these particles in the Galactic center region. These particle distributions are
then used to derive the characteristics of the main nuclear interaction gamma-ray
lines.}
{Because the time period of star capture by the supermassive black hole is
expected to be shorter than the lifetime of the ejected fast
particles against Coulomb losses, the gamma-ray emission is
predicted to be stationary. We find that the nuclear de-excitation
lines should be emitted from a region of maximum 5$^\circ$ angular
radius. The total gamma-ray line flux below 8 MeV is calculated to
be $\approx$10$^{-4}$~photons~cm$^{-2}$~s$^{-1}$. The most
promising lines for detection are those at 4.44 and $\sim$6.2~MeV,
with a predicted flux in each line of
$\approx$$10^{-5}$~photons~cm$^{-2}$~s$^{-1}$. Unfortunately, it
is unlikely that this emission can be detected with the INTEGRAL
observatory. But the predicted line intensities appear to be
within reach of future gamma-ray space instruments. A future
detection of de-excitation gamma-ray lines from the Galactic
center region would provide unique information on the high-energy
processes induced by the central supermassive black hole and
the physical conditions of the emitting region.}
{}

\titlerunning{Nuclear interaction gamma-ray line emission}
\authorrunning{Dogiel et al.}
\maketitle
%

\section{Introduction}

Massive black holes (MBH) at galactic centers are sources of high
energetic activity. X-ray observations of these sources revealed flares
of hard X-ray photons releasing a maximum power of
$10^{44}$~erg~s$^{-1}$. These flares are supposed to be due to
processes of accretion and tidal disruption of stars by the massive
black holes \citep[see, e.g.,][and references thererin]{sirota}.

The capture radius of a black hole - the maximal distance from the
MBH where the tidal forces can overwhelm the stellar
self-gravity and tear the star apart - is given by
\begin{equation}\label{rt}
  R_T \approx 1.4\times 10^{13} \left(\frac{M_{\rm bh}}{10^6M_\odot}\right)^{1/3}
  \left(\frac{M_\ast}{M_\odot}\right)^{-1/3} \left(\frac{R_\ast}{R_\odot}\right)~\mbox{cm}\,,
\end{equation}
where $M_*$ and $R_*$ are the star's mass and radius,
$M_{\rm bh}$ the mass of the black hole, and $M_\odot$ and $R_\odot$
the solar mass and radius.  When a star comes within
the capture radius of a black hole, the tidal force produced by
the black hole is strong enough to capture the star \citep{rees,Phi89}.
If the mass of the black hole is higher than
$\sim$10$^8$~M$_\odot$, the star directly falls into the black hole
event horizon and not much interesting phenomena can be observed.
However, if the mass of the black hole is less than
$\sim$10$^8$~M$_\odot$, the star is torn apart. Roughly half of the
star is captured by the black hole and the disrupted debris
form in a few months a circular transient accretion disk
around the black hole \citep{ulmer}. The time-dependent accretion
rate follows a simple power law, $dM/dt \propto t^{-5/3}$, with an
initial accretion higher than the Eddington rate.

A total tidal disruption of a star occurs when the penetration
parameter $b^{-1}\gg 1$, where $b$ is the ratio of $r_p$ - the
periapse distance (distance of closest approach) to the tidal
radius $R_T$. The tidal disruption rate $\nu_s$ can be
approximated to within an order of magnitude from an analysis of
star dynamics near a black hole via the Fokker-Planck equation.
For the parameters of the GC it gives the rate $\nu_s\sim
10^{-4}$~years$^{-1}$ \citep[see the review of][]{alex05}, which
is in agreement with more detailed calculations of \citet{syer}
who obtained $\nu_s \sim 5\times 10^{-5}$ years$^{-1}$. Tidal
disruption processes were perhaps already observed in cosmological
galaxy surveys \citep[see, e.g.][]{don}.

As observations show, there is a supermassive black hole (Sgr
A$^\ast$) in the center of our Galaxy with a mass of $(4.31 \pm
0.06)\times10^6~M_{\odot}$ \citep{guss}. In the close vicinity of
the Galactic black hole (less than 0.04~pc from it) about 35 low
mass stars (1--3~$M_\odot$) and about 10 massive stars
(3--15~$M_\odot$) are present \citep[see][]{alex},  i.e.,
processes of star accretion in the Galactic center seems to be
quite possible. However, attempts to find a strong X-ray source in
the Galactic Center (GC) were failed. The {\it Chandra} X-ray
Observatory resolved only a weak X-ray point source at the
position of Sgr A$^\ast$ with a flux $L_X\sim
10^{33}$~erg~s$^{-1}$ \citep[see][]{bag}, though moderate X-ray
flares were observed by {\it Chandra} \citep{pouq}. As it was
mentioned by \citet{koyama} and \citet{muno04} this ``X-ray
quiet'' Sgr A$^\ast$ is in sharp contrast to the high X-ray
activity of the surrounding diffuse hot plasmas. One likely
scenario is that the Galactic nucleus was brighter in the past,
possibly caused by a surge accretion onto the massive back hole.

\citet{cheng1,cheng2} suggested that this scenario may explain the
origin of the 511 keV annihilation flux from the GC region, if up
to 10\% of captured stellar matter is ejected in the form of a jet of
relativistic protons. In this case the origin of the 511 keV line emission
from the GC region is supposed to be due to annihilation of
secondary positrons generated by $p-p$ collisions.

An attempt to find independent evidences for active processes in
the GC in the form of a flux of fast charged particles was
undertaken by \citet{dog08}, who concluded that these primary
relativistic protons (if were generated) penetrating into dense
molecular clouds produced there a flux of nuclear de-excitation
gamma-ray lines coming from the GC region. However, the origin of
relativistic protons is still rather speculative since we do not
have direct evidences in favour of their production near black
holes \citep[see, however,][]{auger,ist} . Besides, the gamma-ray
line flux would be strongly time variable in this case and,
therefore, the estimates of its value at present are highly
model-dependent.

Below we examine a different model of gamma-ray line emission,
assuming nuclear interactions of subrelativistic protons generated
by processes of star disruption at the GC.


\section{Flux of subrelativistic protons generated by star disruption}

 An alternative (and to our view more reliable) mechanism of
gamma-ray line production follows from the analysis of processes of
star disruption near black holes. The energy budget of a tidal
disruption event follows from analysis of star matter dynamics.
Once passing the pericenter, the star is tidally disrupted into a
very long and dilute gas stream. The outcome of tidal disruption
is that some energy is extracted out of the orbit to unbind the
star and accelerate the debris. Initially about 50\% of the
stellar mass becomes tightly bound to the black hole , while the
remainder 50\% of the stellar mass is forcefully ejected
\citep[see, e.g.][]{ayal}. The kinetic energy carried by the
ejected debris is a function of the penetration parameter $b^{-1}$ and
can significantly exceed that released by a normal supernova
($\sim 10^{51}$~erg) if the orbit is highly penetrating
\citep[see][]{alex05},
\begin{equation}\label{energy}
  W\sim 4\times 10^{52}\left(\frac{M_\ast}{M_\odot}\right)^2
  \left(\frac{R_\ast}{R_\odot}\right)^{-1}\left(\frac{M_{\rm bh}/M_\ast}{10^6}\right)^{1/3}
  \left(\frac{b}{0.1}\right)^{-2}~\mbox{erg}\,.
\end{equation}
Thus, the mean kinetic energy per escaping nucleon is given by
\begin{equation}\label{esc}
  E_{\rm esc}\sim 42 \left(\frac{\eta}{0.5}\right)^{-1} \left(\frac{M_\ast}{M_\odot}\right)
  \left(\frac{R_\ast}{R_\odot}\right)^{-1}\left(\frac{M_{\rm bh}/M_\ast}{10^6}\right)^{1/3}
  \left(\frac{b}{0.1}\right)^{-2}~\mbox{MeV}\,,
\end{equation}
where $\eta M_\ast$ is the mass of escaping material. For the black-hole mass
$M_{\rm bh}=4.31 \times 10^6~M_{\odot}$ the energy of escaping particles is
$E_{\rm esc} \sim 68 (\eta /0.5)^{-1} (b/0.1)^{-2}~\mbox{MeV nucleon$^{-1}$}$
when a one-solar mass star is captured. The parameters $\eta$ and $b$ are not well
constrained by theory. It is clear, however, that $b$ must be less than unity for a
star to be disrupted in the gravitational potential of the black hole. We take the energy
distribution of the erupted nuclei as a simple Gaussian distribution
\begin{equation}\label{Qesc}
  Q_0^{\rm esc}=\frac{N}{\sigma\sqrt{2\pi}}
  \exp\left[-\,\frac{(E-E_{\rm esc})^2}{2\sigma^2}\right],
\end{equation}
where $N$ is total amount of particles ejected by one stellar
capture. Results of calculations for different values of $\sigma$
do not differ much from each other as long as $\sigma \leq E_{\rm esc}$ (see below).

For a single capture of a one-solar mass star the total number of
unbounded particles is $N \lsim 10^{57}$ and the total kinetic
energy in these particles is $W \lsim 10^{53}$~erg,
if their energy is about  68 MeV (see above).
For the star capture
frequency $\nu_s \simeq 10^{-4}$~year$^{-1}$ and $E_{esc}$ about
several ten MeV (see \cite{dog_pasj}) it gives a power input
$\dot{W} \lsim 3\times10^{41}$~erg~s$^{-1}$, which is a few times
more than the total power contained in Galactic cosmic rays. The
average energy output depends, of course, on the capture
frequency.

A capture of a massive star should eject more fast particles into
the GC region than a capture of a low mass star. However, the
energy of the ejected particles is expected to be similar. This
can be seen from Eq.~(\ref{esc}), recalling that in first
approximation the radius of main-sequence stars varies as $R_\ast
\propto M_\ast^{0.8}$. The frequency of massive star capture is
lower than that for low mass stars \citep[see, e.g.][]{syer}.
Therefore, we expect the capture of a massive star to have
relatively little effect in comparison with the cumulative effect
from low-mass star captures. The latter is calculated below
assuming $M_*=1$~$M_\odot$ for all captured stars.

\section{Proton propagation in the Galactic center region}

 Details of the proton spatial distribution in the GC
region, as well as the mechanism of proton propagation are
inessential for calculations of the total gamma-ray line flux.
We use, nevertheless, kinetic equations in which processes of propagation
are included, and then integrate the obtained solution over the volume
of emission. The effect of propagation reduces in this case to a proper
estimate of the proton flux leaving the emission region.

Propagation of cosmic rays in the Galaxy is described as a diffusion
phenomenon, with a phenomenological diffusion coefficient whose value and
energy dependence is derived from observational data, e.g. from the chemical
composition of cosmic rays measured near Earth. This method leads to a
diffusion coefficient of about $10^{27}$~cm$^2$~s$^{-1}$
\citep[for details see][]{ber90}. Processes of particle propagation in the
GC region are questionable since we do not know much about the physical
conditions of the ambient medium.

From the general theory of cosmic ray origin it is clear that
particle propagation in the interstellar medium is described as
diffusion due to scattering on magnetic fluctuations. The
effective coefficient of diffusion is $D\sim \mathrm{v}^2/\nu(E,
{\bf r})$, where $\mathrm{v}$ is the particle velocity and $\nu(E,
{\bf r})$ is the scattering frequency of particles, which is a
function of particle energy $E$ and coordinates ${\bf r}$. Rather
simple estimates provided by \citet{jean} for the Galactic center
region in the frame of the quasi-linear theory gives $D\sim
10^{27}$~cm$^2$~s$^{-1}$ for MeV particles. A similar estimate was
obtained by \citet{rosa01} from the analysis of radio emission
from the central region. These authors concluded that electrons of
0.14 GeV diffuse about 13 pc in $4\times 10^4$~years, which gives
again $D\sim 10^{27}$~cm$^2$~s$^{-1}$.

Estimates of the diffusion coefficient for subrelativistic protons
in the GC were derived by \citet{dog_pasj2} from the observed hard
X-ray emission, which was assumed to be due to inverse
bremsstrahlung. They estimated $D$ to be in the range $10^{26} -
10^{27}$~cm$^2$~s$^{-1}$. Below we provide our calculations for
the diffusion coefficient $D = 10^{27}$ cm$^2$ s$^{-1}$, though,
in principle, its value may be a function of particle energy (as
well as of spatial coordinates and even time). We suppose,
however, that this simplification is acceptable because the energy
range of relevance for gamma-ray line production is quite narrow,
less than one order of magnitude.

Another important parameter of the model is the mean density of
the medium into which the fast nuclei propagate. Observations show
that the interstellar medium near the GC is highly nonuniform.
Thus, within the central 2~pc of the shell of the supernova
remnant Sgr~A East, there is a hot plasma of temperature 2~keV and
density as high as $10^3$~cm$^{-3}$ \citep{maeda}. This local
density fluctuation is insignificant for our treatments since
energetic particles capable of producing de-excitation gamma-ray
lines lose little energy passing through this local fluctuation by
diffusion.

A significant part of the gas in the GC region is in the form of molecular
hydrogen, $\sim$90\% of which is contained in dense and massive molecular
clouds with a volume filling factor of only a few per cent. The total mass
of molecular hydrogen in the Nuclear Bulge, which is estimated to be
$2\times 10^7~M_\odot$, significantly exceeds the mass contained in
the intercloud hot plasma \citep[see][]{laun}.

However, subrelativistic charged particles may not be able to
penetrate freely inside molecular clouds \citep[see,
e.g.,][]{skill}. Theoretical investigations performed by Dogiel et
al. (1987, 2005) showed that turbulent motions of neutral gas
could excite small scale electromagnetic fluctuations that could
prevent charged particles to penetrate deeply into clouds. Thus,
subrelativistic cosmic-ray particles are expected to fill only a
small portion of molecular clouds. This conclusion is supported by
calculations performed by \citet{dog_pasj1}, who showed that the
flux of bremsstrahlung emission produced by subrelativistic
protons inside molecular clouds is lower than that produced in the
intercloud hot plasma. If this is true, we expect that most of the
gamma-ray line emission should also be generated in the intercloud
hot plasma.

Recent observations with the satellites {\it ASCA}, {\it Chandra}
and {\it Suzaku} showed that the $1-2^\circ$ radius central region
is filled with a hot gas of temperature $6 - 10$ keV.
The density of plasma derived from these observations ranges between
$n\simeq 0.1-0.4$ cm$^{-3}$ \citep{koyama,muno04,koyama07}.
In the calculations we use the average value $n= 0.2$ cm$^{-3}$.
For these parameters the
average time of Coulomb losses for 10--100 MeV protons is
\citep[see, e.g.][]{haya}
\begin{equation}
  \tau_i\sim \frac{E_pm_e\mathrm{v}_p}{6\pi ne^4\ln\Lambda}\sim
  3\cdot 10^{6} - 3\cdot 10^{7}~\mbox{yr}\,,
  \label{taui}
\end{equation}
where $E_p$ and $\mathrm{v}_p$ are the proton energy and velocity,
and $\ln\Lambda$ is the Coulomb logarithm.

\section{Spatial and energy distributions of subrelativistic protons
in the Galactic center region}

\begin{figure}
\centering \includegraphics[width=9cm]{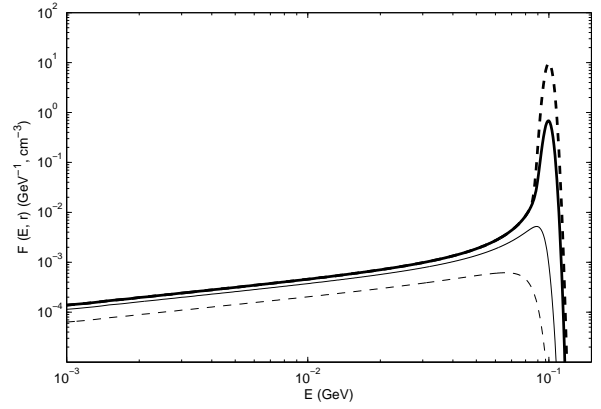}
\caption{Energy spectra of primary protons at $r=0$~pc and
$t=10^3$~years ({\it thick dashed line}), $r=0$~pc and
$t=10^4$~years ({\it thick solid line}), $r=150$~pc and
$t=10^4$~years ({\it thin solid line}), and $r=300$~pc,
$t=10^4$~years ({\it thin dashed line}). For $r>10$~pc the proton
spectra are almost stationary. These spectra were calculated for
$E_{\rm esc}=100$~MeV and $\sigma=0.05 E_{\rm esc}$. }
\label{E_dist}
\end{figure}

The time-dependent spectrum of subrelativistic protons
erupted from the central region can be calculated from the equation
\begin{equation}\label{pr_state}
  \frac{\partial f}{\partial t}  - \nabla \left(D\nabla f \right)+
  \frac{\partial}{\partial E}\left( \frac{dE}{dt} f\right) =
  Q(E,t)\,,
\end{equation}
where $dE/dt\equiv b(E)$ is the rate of  proton energy losses.
Subrelativistic protons lose their energy by Coulomb collisions.
The rate of losses for these protons can be presented in the form
\begin{equation}\label{ion_nr}
  b(E)=-\frac{2\pi ne^4}{m_e \mathrm{v}}\ln\Lambda\simeq-\frac{a}{\sqrt{E}}\,.
\end{equation}
The injection of protons by processes of star capture can be
described by
\begin{equation}
  Q(E, {\bf r}, t) = \sum \limits_{k=0}Q_k(E)\delta(t - t_k)\delta({\bf r})\,,
\end{equation}
where $t_k=k \times T$ is the injection time and the functions $Q_k(E)$
are given by Eq.~ (\ref{Qesc}). The mean time of star capture by the
massive black hole in the Galaxy is taken to be $T\simeq 10^4$~years.

We can derive the Green function of Eq.~(\ref{pr_state}) for the
injection time $t_k$ \citep[see][]{syr,grat} and then
sum over injections. The equation for the Green function is
\begin{equation}
  {{\partial G_k}\over{\partial t}}-\nabla \left(D\nabla G_k\right)
  +{\partial\over{\partial E}} \left(b(E)G_k\right)=\delta(E-E_0)
  \delta(t-t_k)\delta({\bf r})\,.
\end{equation}
Using variables
\begin{equation}
  \tau(E,E_0)=\int\limits^E_{E_0}{{dE}\over{b(E)}}~~~~~\mbox{and}~~~~~
  \lambda=\int\limits^E_{E_0}{{D(E)}\over{b(E)}}dE
\end{equation}
the Green function can be written as
\begin{equation}
  G_k({\bf r},E,t;E_0,t_k)={{1}\over{\mid b(E)\mid}}
  {{\delta(t-t_k-\tau)}
  \over{(4\pi\lambda)^{3/2}}}\exp\left[-\frac{{\bf r}^2}{4\lambda}\right]
\end{equation}
Then $f({\bf r},E,t)$ is
\begin{equation}\label {sol11}
  f({\bf r},E,t)=\sum\limits_{k=0}\int_0^\infty dE_0
  Q_k(E_0)G_k({\bf r},E,t;E_0, t_k)
\end{equation}

In the nonrelativistic case, i.e. for energy losses in the form
of Eq.~(\ref{ion_nr}), the solution (\ref{sol11}) can be simplified to
\begin{equation}\label{sol1}
  f({\bf r},E,t)=\sum\limits_{k=0}\frac{N_k\sqrt{E}}{\sigma\sqrt{2\pi}Y_k^{1/3}}
  \frac{\exp\left[-\frac{\left(E_{\rm esc}-Y_k^{2/3}\right)^2}{2\sigma^2}-
  \frac{{\bf r}^2}{4D(t-t_k)}\right]}{ \left(4\pi D(t-t_k)\right)^{3/2}}\,,
\end{equation}
where
\begin{equation}
  Y_k(t,E)=\left[\frac{3a}{2}(t-t_k)+E^{3/2}\right]\,.
\end{equation}

\begin{figure}
\centering \includegraphics[width=9cm]{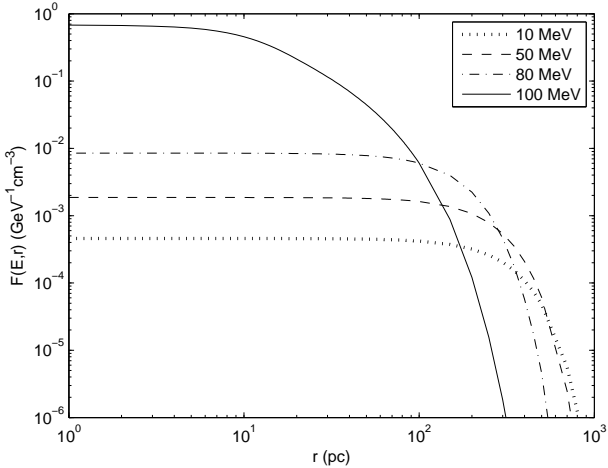}
\caption{Densities in protons of energies 10, 50, 80 and 100~MeV, as a
function of radial distance from the GC. These spatial
distributions were calculated for $E_{\rm esc}=100$~MeV,
$\sigma=0.05 E_{\rm esc}$,
and $n=0.2$ cm$^{-3}$.
}\label{r_dist}
\end{figure}

As one can see this solution is characterized by two different
spatial scales. The first one is $R_1\sim\sqrt{4DT}\sim 10$ pc.
Inside the sphere $r<R_1$ the solution is time dependent and the
density of protons is strongly fluctuating with time. On the other
hand, for $r>R_1$ the solution is quasi-stationary because the
time of diffusion over these distances is larger than the mean
capture time $T$. The radius of the sphere filled with
subrelativistic protons (the scale of exponential decrease) is
about $R_2\sim\sqrt{4D\tau_i}\sim 200$~pc for $E_{\rm
esc}=100$~MeV.
One can see from Eq. (\ref{taui}) that the mean free path
of protons is proportional to $\sqrt{1/n}$.
 Examples of calculated energy and spatial distributions of
subrelativistic protons are shown in Figs. \ref{E_dist} and
\ref{r_dist}.

Since the Galactic black hole was inactive for a long time, we
use to calculate the gamma-ray line emission the proton spectrum at
time $t=T$ after the last star capture. This will give us a lower limit
on the gamma-ray line flux. However, the density of subrelativistic
protons in the GC is expected to be almost independent of time, except
for a very compact region around the massive black hole. The proton
spectrum should reach its saturation level after $\gsim \tau_i/T$ capture
events. In our calculations we summarized the cumulative effect of 2000
captures.

In order to predict the total gamma-ray line flux from the GC we need
to estimate the total (integrated over volume) spectrum of subrelativistic
protons
\begin{equation}
F(E,t)=4\pi\int\limits_0^\infty f({\bf r},E,t)r^2dr. \label{exact}
\end{equation}

We note that for simple estimates one can derive this spectrum from
the assumption of a stationary injection of protons with the rate
$Q=N/T$. Then the approximate solution has the form
\begin{equation}\label{approx}
  \bar{F}(E)=\frac{Q\sqrt{E}}{2a}\left[1+\mbox{Erf}\left(\frac{E_{\rm esc}-E}
  {\sqrt{2}\sigma}\right)\right]
\end{equation}
where Erf$(x)$ is the error function.

\section{Nuclear interaction gamma-ray line emission from the Galactic
center region}

Unlike relativistic particles, subrelativistic nuclei do not
produce effectively continuous radiation (except, may be, inverse
bremsstrahlung, which is discussed in Dogiel et al.
2009abc). However, collisions of subrelativistic nuclei with
ambient matter can lead to nuclear excitation and result in emission
of de-excitation gamma-ray lines. These lines may be a good tracer
for subrelativistic cosmic rays, because the line brightness can
give us information about the amount of subrelativistic
particles.

Thus, prominent nuclear de-excitation lines are observed in
emission spectra of solar flares, which allows one to derive
valuable information on solar ambient abundances, density and
temperature, as well as on accelerated particle composition,
spectra and transport in the solar atmosphere (see e.g.,
\citet{smith} and \citet{kiener} for recent solar observations
with {\it RHESSI} and {\it INTEGRAL}, respectively). \citet{dog01}
predicted that galaxy clusters could emit a detectable flux of
de-excitation gamma-ray lines. Latter, \citet{iyud} found traces
of gamma-ray line emission towards the Coma and Virgo clusters,
though this detection has not been confirmed yet.

Detection of the gamma-ray line emission produced by cosmic-ray
interactions in the interstellar medium would provide insightful
information on low-energy cosmic rays and give a significant
advance in the development of the theory of cosmic-ray origin.
Theoretical estimates for the line emission in the Galaxy were
provided by \citet{ram79}, who calculated spectra of the nuclear
de-excitation line emission for different assumed spectra of
subrelativistic ions. However, attempts to measure a prominent
flux of nuclear interaction lines were failed up to now.

Galactic center observations in the MeV range were performed by
the COMPTEL group but with a rather poor angular resolution
\citep{bloeb}. They detected a marginal excess of emission in the
range 3--7 MeV. The total flux from the central region
$60^{\circ}\times 20^{\circ}$ was estimated to be
$10^{-4}$~cm$^{-2}$~s$^{-1}$~rad$^{-1}$. This excess was
interpreted as emission in nuclear de-excitation lines generated
by cosmic-ray interaction in the GC region. However, neither the
COMPTEL group nor subsequent observations with {\it INTEGRAL} have
confirmed this result.  In particular, \citet{tee06} have carried
out an extensive search for gamma-ray lines in the first year of
public data from the {\it INTEGRAL} spectrometer SPI. They found
no evidence for any previously unknown lines in the energy range
20--8000~keV.

We suppose that processes of star accretion by the massive black hole
at the GC can generate a significant number of subrelativistic
protons and heavier nuclei. These energetic particles could in turn
produce a significant flux of nuclear de-excitation gamma-ray lines in a
relatively compact region of the GC. Unlike the gamma-ray
emission that might be produced by relativistic protons \citep{dog08},
the emission produced by subrelativistic protons is expected to be almost
stationary and should therefore be less dependent on model parameters.

Though according to the present model the origin of
subrelativistic cosmic rays near the GC is different
than in other part of the galactic disk, a positive detection of
de-excitation gamma-ray lines would provide important information
about star accretion processes at the GC.

\begin{figure}
\centering \includegraphics[width=7.5cm]{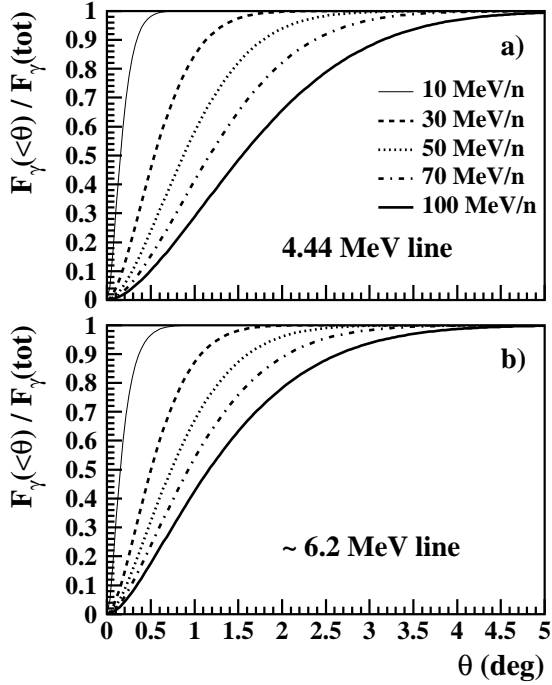}
\caption{Radial profiles of enclosed fluxes in (a) the 4.44~MeV line and
(b) the $\sim$6.2~MeV line complex, as a function of angular radius
from the GC, for the five values of $E_{\rm esc}$ indicated
in the figure. The calculations assume $\sigma=0.05 E_{\rm esc}$ (Eq.~\ref{Qesc})
and $n_{\rm H}=0.2$~cm$^{-3}$.
}\label{gamspatial}
\end{figure}

We calculate the gamma-ray flux emitted at time $t=T$ after a star
capture in a given nuclear de-excitation line from
\begin{equation}\label{fluxgam}
  F_\gamma(t)={n_H \over 4 \pi d^2} \sum_{ij} {n_j \over n_H}
  \int_0^\infty F_i(E_i,t) \mbox{v}_i(E_i) \sigma_{ij}(E_i) dE_i~,
\end{equation}
where $i$ and $j$ range over the accelerated and ambient particle
species that contribute to the production of the gamma-ray line
considered, $n_{\rm H}=0.2$~cm$^{-3}$ is the adopted mean density
of H atoms in the interaction region, $n_j$ is the density of the
ambient constituent $j$, $F_i(E_i,t)$ is the energy spectrum at
time $t$ of the fast particles in the interaction region (Eq.~15),
v$_i$ is the velocity of these particles, $\sigma_{ij}$ is the
cross section for the reaction of interest between species $i$ and
$j$, and $d=8$~kpc is the distance to the Galactic center
\citep{gro08}. To take into account the measured enhancement of
the metal abundances in the Galactic center region \citep{cun07},
we assume the abundances of ambient carbon and heavier elements to
be twice solar \citep{lod03}. This is consistent with the recent
estimate from {\it Suzaku} observations of \citet{koy08}, who
found the Fe abundance to be between 1.5 and 2.3 times solar. In
the case of a low-mass star capture, we expect the line emission
to be mainly due to proton and $\alpha$-particle reactions with
ambient heavy nuclei. We took into account the $\alpha$ particles
assuming, for simplicity, $F_\alpha(E,t)=X_\alpha F_p(E,t)$, where
the energy $E$ is expressed in MeV~nucleon$^{-1}$ and
$X_\alpha=0.1$ is the accelerated $\alpha/p$ abundance ratio.

\begin{figure}
\centering \includegraphics[width=7.5cm]{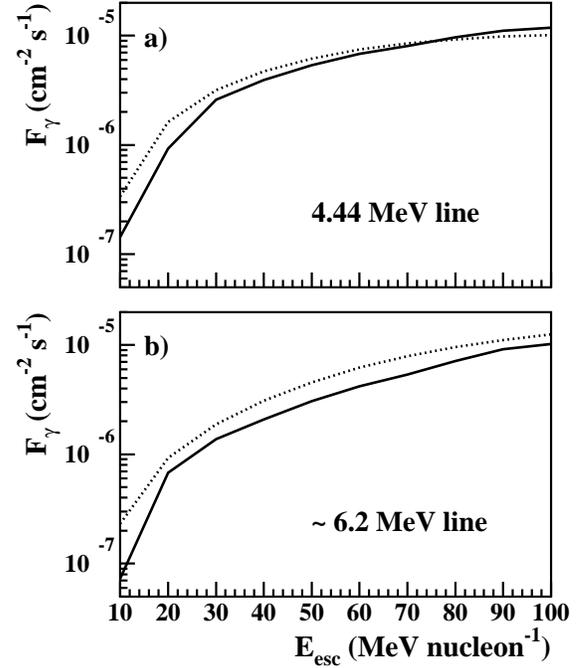}
\caption{Predicted fluxes from the GC region of
angular radius $\theta=5^\circ$ in (a) the 4.44~MeV line and (b) the
$\sim$6.2~MeV line complex, as a function of $E_{\rm esc}$, for
$\sigma=0.05 E_{\rm esc}$ ({\it solid lines}) and $\sigma=E_{\rm esc}$
({\it dashed lines}).
}\label{gamen}
\end{figure}

 Noteworthy, the total steady-state emission obtained from Eq.~(\ref{fluxgam})
is independent of the assumed mean density in the interaction region, as can be seen
by introducing in this equation the particle distribution $F_i(E,t)$, which is
inversely proportional to $n_{\rm H}$ (see, e.g., Eq.~\ref{approx}). This is true,
however, as long as $n_{\rm H} \lsim 100$~cm$^{-3}$, because for larger density
values the Coulomb loss time $\tau_i < T$ and the particle distribution is not
anymore stationary. In any case, even if the predicted emission were time dependent,
the total fluxes given below can be interpreted as mean values over the period
of star capture $T$.

Figure~\ref{gamspatial} shows calculated radial profiles of the predicted line
emission at 4.44 and $\sim$6.2~MeV. These two lines are mainly produced by proton
and $\alpha$-particle reactions with ambient $^{12}$C and $^{16}$O (see below).
Contrary to the total emission from the GC region (i.e. integrated over space), the
radial profiles depend on the assumed value of $n_{\rm H}$: the larger the ambient
medium density the lower the spatial extend of the gamma-ray line emission. We see in
Fig.~\ref{gamspatial} that for $n_{\rm H}=0.2$~cm$^{-3}$ and
$E_{\rm esc} \leq 100$~MeV~nucleon$^{-1}$, the gamma-ray lines are emitted from a
region of maximum $5^\circ$ angular radius. Noteworthy, such an emission would
appear as a small scale diffuse emission for a gamma-ray instrument like SPI, whose
angular resolution is $\sim$3$^{\circ}$ FWHM \citep[see, e.g.,][]{tee06}.

Figure~\ref{gamen} shows calculated fluxes in the 4.44 and $\sim$6.2~MeV lines as a
function of $E_{\rm esc}$ ranging from 10 to 100~MeV~nucleon$^{-1}$, which corresponds
to $0.68 < (\eta /0.5) (b/0.1)^2 < 6.8$ (see Eq.~\ref{esc}).
We see that the predicted fluxes increase with increasing escape energy up to values
$\approx 10^{-5}$~photons~cm$^{-2}$~s$^{-1}$ for $E_{\rm esc} \approx 100$~MeV~nucleon$^{-1}$.
We also see that the fluxes are not strongly dependent of the assumed
width of the adopted Gaussian distribution ($\sigma$ in Eq.~4).

\begin{figure}
\centering \includegraphics[width=9.cm]{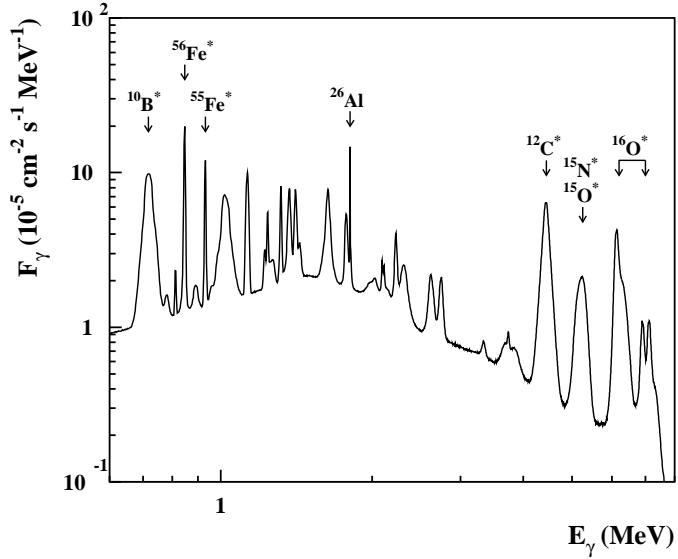}
\caption{Spectrum of the predicted gamma-ray line emission from the GC region
of angular radius $\theta=5^\circ$, assuming
$E_{\rm esc}=100$~MeV~nucleon$^{-1}$ and $\sigma=0.05 E_{\rm esc}$. The arrows
point to the lines which are mentioned in the text.
}\label{gamma}
\end{figure}

\citet{dog_pasj} showed that the hard X-rays observed with {\it
Suzaku} from the GC region \citep{yuasa} could be produced by
inverse bremsstrahlung from subrelativistic protons accelerated by
star disruption near Sgr~A$^\ast$. In this scenario, the measured
spectrum of this hard X-ray emission would imply that
$E_{esc}\approx 100$~MeV~nucleon$^{-1}$. Figure~\ref{gamma} shows
a gamma-ray line spectrum calculated for
$E_{esc}=100$~MeV~nucleon$^{-1}$ and $\sigma=0.05 E_{\rm esc}$,
using the code developed by \citet{ram79}. The calculated total
flux in gamma-ray lines below 8~MeV is
$1.1\times10^{-4}$~photons~cm$^{-2}$~s$^{-1}$. In the 3--7~MeV
range, it is $4.4\times10^{-5}$~photons~cm$^{-2}$~s$^{-1}$.

We see in Fig.~\ref{gamma} relatively strong lines of various
widths, which are superimposed on a continuum-like emission due to
a large number of unresolved gamma-ray lines, mostly arising from
cascade transitions in high-lying levels of heavy nuclei. Among
the strongest lines are those produced in excitations of low-lying
nuclear levels of abundant ambient ions, e.g. at 4.44~MeV from
$^{12}$C$^*$ and 0.847~MeV from $^{56}$Fe$^*$. Other important
lines are those arising from transitions in the spallation
products of abundant ambient nuclei, e.g. at 0.718~MeV from
$^{10}$B$^*$, which is produced by spallation of ambient $^{12}$C
and $^{16}$O, at 0.931~MeV from $^{55}$Fe$^*$, which is mainly
produced by the reaction $^{56}$Fe($p$,$p$$n$)$^{55}$Fe$^*$, and
at $\sim$5.2~MeV from $^{15}$N$^*$ and $^{15}$O$^*$ resulting from
spallation of $^{16}$O \citep[see][]{koz02}. The broad line
feature at $\sim$6.2~MeV is due to the merging of the 6.129~MeV
line from de-excitations in $^{16}$O$^*$ with two lines at
6.175~MeV from $^{15}$O$^*$ and 6.322~MeV from $^{15}$N$^*$
produced in spallation reactions.  The characteristics of the main
gamma-ray lines are summarized in Table~\ref{table1}.

The lines produced in a gaseous ambient medium are broadened by
the recoil velocity of the excited nucleus. The line width
generally decreases with increasing mass of the target atoms,
because the recoil velocity of heavier nuclei is smaller. Thus, in
Fig.~\ref{gamma}, the full width at half maximum (FWHM) of the
0.847~MeV line produced by proton and $\alpha$-particle inelastic
scattering of $^{56}$Fe is 6~keV (0.7\% of the transition energy),
whereas the FWHM of the 4.44~MeV from $^{12}$C$^*$ is 160~keV
($\Delta E_\gamma / E_\gamma=3.6$\%).

\begin{table}
\caption{Main gamma-ray lines expected from the GC region}
\label{table1}
\centering
\begin{minipage}{0.45\textwidth}
\begin{tabular}{ c c c c }
\hline
\hline
Line energy & Nuclear    & FWHM$^a$ & Flux$^a$ \\
(MeV)       & Transition & (keV)                & cm$^{-2}$ s$^{-1}$   \\
\hline
0.718 & $^{10}$B$^*_{0.718}$ $\rightarrow$ g.s.$^b$  & 30 & 3.4$\times$10$^{-6}$ \\
0.847 & $^{56}$Fe$^*_{0.847}$ $\rightarrow$ g.s. & 6 &  1.4$\times$10$^{-6}$ \\
      & $^{27}$Al$^*_{0.844}$ $\rightarrow$ g.s. &   &   \\
1.809 & $^{26}$Mg$^*_{1.809}$ $\rightarrow$ g.s. & $<$2.7$^c$ & 3.5$\times$10$^{-7}$ \\
      & ($^{26}$Al decay)                        &   &   \\
4.44  & $^{12}$C$^*_{4.439}$ $\rightarrow$ g.s.  & 160 & 1.2$\times$10$^{-5}$ \\
      & $^{11}$B$^*_{4.445}$ $\rightarrow$ g.s.  &   &   \\
$\sim$5.2 & $^{14}$N$^*_{5.106}$ $\rightarrow$ g.s. & 280 & 5.4$\times$10$^{-6}$ \\
          & $^{15}$O$^*_{5.181}$ $\rightarrow$ g.s. &     &   \\
          & $^{15}$O$^*_{5.241}$ $\rightarrow$ g.s. &     &   \\
          & $^{15}$N$^*_{5.270}$ $\rightarrow$ g.s. &     &   \\
          & $^{15}$N$^*_{5.299}$ $\rightarrow$ g.s. &     &   \\
$\sim$6.2 & $^{16}$O$^*_{6.130}$ $\rightarrow$ g.s. & 180$^d$ & 1.0$\times$10$^{-5}$ \\
          & $^{15}$O$^*_{6.176}$ $\rightarrow$ g.s. &   &   \\
          & $^{15}$N$^*_{6.324}$ $\rightarrow$ g.s. &   &   \\
\hline
\end{tabular}
\footnotetext[1]{For $E_{\rm esc}$=100 MeV nucleon$^{-1}$ and $\sigma$=0.05$E_{\rm esc}$
(see Eq.~4). The fluxes are for the region of angular radius $\theta=5^\circ$ (see
Fig.~\ref{gamspatial} for fluxes at 4.44 and $\sim$6.2~MeV calculated for lower values of
$\theta$).}
\footnotetext[2]{Ground state.}
\footnotetext[3]{Assuming thermal broadening with $T<10$~keV.}
\footnotetext[4]{A very narrow line component at 6.129~MeV could arise from excitation
of $^{16}$O nuclei contained in micrometer-sized grains (see text).}
\end{minipage}
\end{table}

However, some gamma-ray lines produced in interstellar dust grains
can be very narrow, because some of the excited nuclei can stop in
solid materials before emitting gamma-rays \citep{lin77}. It
requires that the mean lifetime of the excited nuclear level or of
its radioactive nuclear parent is longer than the slowing down
time of the excited nucleus in the grain material. The most
promising of such lines is that at 6.129~MeV from $^{16}$O$^*$.
Thus, future measurements of the profile of the $\sim$6.2~MeV line
could allow us to probe the presence of dust grains in the
Galactic center region and perhaps to
measure the grain size distribution \citep{tat04}.

Very narrow lines can also result from the decay of
spallation-produced long-lived radionuclei, which can come
essentially to rest in the ambient medium before decaying to
excited states of their daughter nuclei and emitting gamma-rays.
For $n_{\rm H} \approx 0.2$~cm$^{-3}$, the slowing-down time of
recoiling heavy nuclei excited by fast proton and $\alpha$-particle
collisions is of the order of $10^4$ years. Thus, only radionuclei
with mean lifetimes longer than $10^4$~yr can produce very
narrow lines in the GC environment. The most intense of
these lines is expected to be that at 1.809~MeV from the decay of
$^{26}$Al ($T_{1/2}=7.2\times10^5$~years). This radioisotope can be
synthesized by the reactions $^{26}$Mg($p$,$n$)$^{26}$Al and
$^{28}$Si($p$,$x$)$^{26}$Al \citep{ram79}. However, the calculated
flux in the 1.809~MeV line is only
3.5$\times$10$^{-7}$~photons~cm$^{-2}$~s$^{-1}$ (Table~\ref{table1}).
In comparison, the flux observed with {\it INTEGRAL}/SPI from the inner
Galaxy ($- 30^\circ < l < 30^\circ;~ -10^\circ < b < 10^\circ$)
is $(3.3\pm0.4)\times 10^{-4}$~cm$^{-2}$~s$^{-1}$~rad$^{-1}$. It
is produced by a present-day equilibrium mass of $2.8 \pm
0.8$~M$_\odot$ of $^{26}$Al synthesized by massive stars in the
Milky Way \citep{diehl}.

 The gamma-ray lines above 4 MeV are better candidates for a
future detection. Our predicted fluxes, however, are below currently
available sensitivity limits of {\it INTEGRAL}. After one year of
observation and $\sim$3~Ms of exposure of the GC region, the SPI
sensitivity for detection of a narrow line at $\sim$5~MeV from a
small-scale diffuse region at the center of the Galaxy was
3.2$\times$10$^{-5}$~cm$^{-2}$~s$^{-1}$ \citep[][see Table~5]{tee06}.
But for a broad line of $\sim$200~keV FWHM (see Table~\ref{table1})
the sensitivity limit increases by a factor of $\sim$8. It is thus
unlikely that the predicted gamma-ray line emission will be detected
with {\it INTEGRAL}.

\section{Discussion and conclusions}
In spite of many uncertainties of the model parameters we
should state that their estimates cannot vary more than an order of magnitude.
Thus, the injection energy should be close to 100 MeV, the average
energy output is about $10^{41}$ erg s$^{-1}$ and the spatial
diffusion coefficient is about $10^{27}$ cm$^2$s$^{-1}$
in order to reproduce the observed spatial
distribution of hard X-rays in the GC
(Dogiel et al. 2009a,b,c).

We showed that accretion of stars onto the central supermassive
black hole periodically eject in the GC region an intense flux of
subrelativistic protons and nuclei capable of producing a significant
emission of nuclear de-excitation gamma-ray lines. The production rate of
these energetic particles is of the order of $10^{45}$~s$^{-1}$. Because
the time period of star capture by the black hole is expected to be shorter
than the lifetime of the ejected protons against Coulomb losses, the density
of these particles in the GC region is expected to be almost stationary. The
radius of the volume filled with these energetic particles is found
to be $\lsim$700~pc or $\lsim$5$^\circ$ if observed from Earth.

Based on the spectrum of the diffuse hard X-ray emission observed with
{\it Suzaku} from the GC region \citep{yuasa}, we estimated the energy of
the nuclei accelerated in the vicinity of the black hole to be
$E_{esc}\approx 100$~MeV~nucleon$^{-1}$. Further assuming that the mean
metallicity in the GC region is two times higher than in the solar
neighborhood \citep[see, e.g.,][]{koy08}, we calculated the total gamma-ray
line flux below 8 MeV to be $1.1\times10^{-4}$~photons~cm$^{-2}$~s$^{-1}$.
The most promising lines for detection are those at 4.44 and $\sim$6.2~MeV,
with a predicted flux in each line of $\approx$$10^{-5}$~photons~cm$^{-2}$~s$^{-1}$.
These lines should be broad, $\Delta E_\gamma / E_\gamma$ of 3--4\%, which
unfortunately renders their detection with the {\it INTEGRAL} spectrometer unlikely.

But future gamma-ray missions like the Nuclear Compton Telescope \citep{chang}, the
GRIPS project \citep{grein}, and the Advanced Compton Telescope \citep{boggs} may be
able to test our predictions. In particular, the GRIPS mission proposed for ESA's
"Cosmic Vision" program could achieve after 5 years in orbit more than an order of
magnitude sensitivity improvement over COMPTEL (in 9 years), which would allow a clear
detection of the predicted gamma-ray line emission at 4.44 and $\sim$6.2~MeV from the GC
region. The Advanced Compton Telescope project proposed as a future NASA mission aims at
even better sensitivity, near 10$^{-6}$~photons~cm$^{-2}$~s$^{-1}$ for 3\% broad lines.
A future detection of the predicted gamma-ray lines with such an instrument would
provide unique information on the high-energy processes induced by the the central
black hole, as well as on the physical conditions of the emitting region.

 Finally, we note that nuclear interactions of subrelativistic ions with ambient
material can also synthesize $\beta^+$ radioisotopes, whose decay can inject positrons
into the GC region. From the radioisotope production yields given by \citet{koz87}, we
estimate that for $E_{esc}\approx 100$~MeV~nucleon$^{-1}$ the number of positrons
produced by this mechanism is $\lsim$5 times the number of gamma-rays emitted in the
4.44 and $\sim$6.2 MeV lines. With a predicted gamma-ray production rate of
$1.7 \times 10^{41}$~photons~s$^{-1}$ in the sum of these two lines (see Table~1),
it gives a positron production rate $\lsim$$8.4 \times 10^{41}$~$\beta^+$~s$^{-1}$.
This limit is more than an order of magnitude lower than the positron annihilation
rate measured with {\it INTEGRAL}/SPI \citep{wei08}.



\begin{acknowledgements}
 First of all we would like to mention that the referee's
report was very helpful for us and we thank him for comments. The
authors are also grateful to H.-K. Chang, Y. Chou, Ya. N. Istomin,
and Y. Maeda,  for helpful discussions, and to W. Hermsen and H.
Bloemen for valuable inputs.

VAD and DOC are partly supported by the RFBR grant 08-02-00170-a,
the NSC-RFBR Joint Research Project RP09N04 and
09-02-92000-HHC-a and by the grant of a President of the Russian
Federation "Scientific School of Academician V.L.Ginzburg".
KSC is supported by a RGC grant of Hong Kong Government under HKU 7014/07P
and a National Basic Research Program of China under 2009CB824800.
CMK is supported by the Taiwan National Science Council grants NSC
96-2112-M-008-014-MY3 and NSC-98-2923-M-008-001-MY3.
WHI is supported by the Taiwan National Science Council grants NSC
96-2752-M-008-011-PAE and NSC 96-2111-M-008-010.

The last version of the paper was partly prepared in ISAS/JAXA
(Sagamihara, Japan). VAD thanks ISAS and particularly Prof. Kazuhisa
Mitsuda and the ISAS director Prof. Hajime Inoue for hospitality.
\end{acknowledgements}

\end{document}